\documentclass[prd,aps,floats,twocolumn,showpacs]{revtex4}
\usepackage[dvips]{epsfig}

\begin{document}

\def\sqr#1#2{{\vcenter{\hrule height.#2pt
   \hbox{\vrule width.#2pt height#1pt \kern#1pt
      \vrule width.#2pt}
   \hrule height.#2pt}}}
\def\square{{\mathchoice\sqr64\sqr64\sqr{3.0}3\sqr{3.0}3}}

\title{Deconfined SU(2) phase with a massive vector boson triplet}

\author{Bernd A. Berg}

\affiliation{Department of Physics, Florida State
             University, Tallahassee, FL 32306-4350, USA}

\date{\today } 

\begin{abstract}
We introduce a model of SU(2) and U(1) vector fields with a local U(2)
symmetry. Its action can be obtained in the London limit of a gauge 
invariant regularization involving two scalar fields. Evidence from
lattice simulations of the model supports a (zero temperature) SU(2) 
deconfining phase transition through breaking of the SU(2) center 
symmetry, and a massive vector boson triplet is found in the deconfined 
phase.
\end{abstract}
\pacs{11.15.Ha, 12.15.-y, 12.60.Cn, 12.60.-i, 14.80.Bn}
\maketitle

\section{Introduction}

Euclidean field theory notation is used throughout this paper. The 
action of the electroweak gauge part of the standard model reads
\begin{eqnarray} \label{Sew}
  S &=& \int d^4x\,L^{\rm ew}\,,\\ \label{Lew}
  L^{\rm ew} &=& - \frac{1}{4}\,F^{\rm em}_{\mu\nu} 
  F^{\rm em}_{\mu\nu} - \frac{1}{2}\,{\rm Tr}\,
  F^b_{\mu\nu}F^b_{\mu\nu}\,,\\ \label{amu} 
  F^{\rm em}_{\mu\nu}&=&\partial_{\mu}a_{\nu}-\partial_{\nu}a_{\mu}\,,
  \label{F1} \\ F^b_{\mu\nu}&=&\partial_{\mu}B_{\nu}-\partial_{\nu}
  B_{\mu} + ig_b \left[B_{\mu},B_{\nu}\right]\,,\label{F2}
\end{eqnarray}
where $a'_{\mu}$ are U(1) and $B_{\mu}$ are SU(2) gauge fields.

Typical textbook introductions of the standard model, e.g.~\cite{Qu83},
emphasize at this point that the theory contains four massless gauge
bosons and introduce the Higgs mechanism so that one obtains a massive
vector boson triplet and only one gauge boson, the photon, stays 
massless. Such presentations reflect that the introduction of the 
Higgs particle in electroweak interactions \cite{We67} preceded our 
non-perturbative understanding of non-Abelian gauge theories. 

In fact, massless gluons are not in the physical spectrum of 
(\ref{Sew}). The self interaction due to the commutator (\ref{F2}) 
generates dynamically a non-perturbative mass gap, and the SU(2) 
spectrum consists of massive glueballs. The lightest glueball can 
be used to set a mass scale. Choosing for it, e.g., 80 GeV and coupling 
fermions is perfectly admissible. This does not constitute an ansatz 
for an electroweak theory by two reasons: The SU(2) glueball spectrum 
is not what is wanted (e.g., masses of spin 0 and 2 states are lower 
than for spin 1 \cite{SU2g}) and fermions would be confined into 
boundstates, which is not the case. Coupling a Higgs field causes a 
deconfining phase transition, so that fermions are liberated, a photon 
stays massless and glueballs break up into elementary massive vector 
bosons. Such a confinement-Higgs transition has indeed been observed 
in pioneering lattice gauge theory (LGT) investigations~\cite{Early}. 
The present paper introduces a different model for which a 
zero-temperature deconfined phase breaks the Z$_2$ center symmetry 
of SU(2) and exhibits a massive vector boson triplet.

\begin{figure}[tb] \begin{center} 
\epsfig{figure=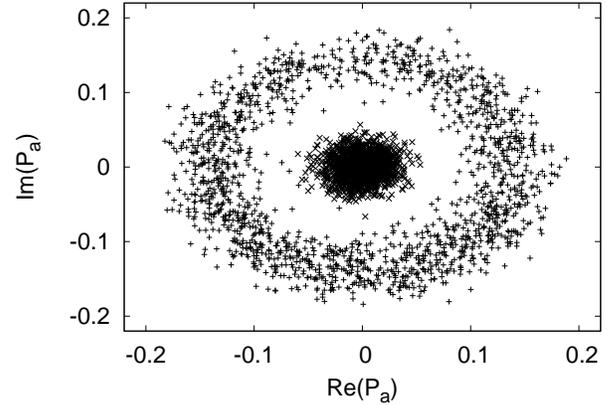,width=\columnwidth} 
\caption{Scatter plot for U(1) Polyakov loops on a $12^4$ lattice 
at $\beta_a=0.9$ in the symmetric phase (center) and at $\beta_a=1.1$ 
in the broken phase (ring). \vspace{-4mm}
\label{fig_U1scatter1} } \end{center} \end{figure} 

The motivation for the action defined in the next section comes from
properties of the U(1) Polyakov loop. U(1) LGT confines fermions in 
the strong coupling limit of its lattice regularization \cite{Wi74}. 
At weaker coupling it undergoes a transition into the Coulomb phase 
as has been rigorously proven \cite{Gu80}. In the Coulomb phase the 
effective potential of the U(1) Polyakov loop $P_a$ assumes the Mexican 
hat shape that is also characteristic for a complex Higgs field in the 
broken phase. This is illustrated in Fig.~\ref{fig_U1scatter1} by 
scatter plots of Polyakov loops in the symmetric and in the broken 
phase. From the ring-like distribution in the broken phase the 
similarity with the behavior of a Higgs field is evident (details 
of the simulations for the figure are given in section~\ref{sec_Num}).

We introduce a contribution to the action that is the alternating 
product of SU(2) and U(1) gauge matrices around a plaquette. It 
unifies SU(2) and U(1) gauge transformations into a local U(2) 
symmetry, which we take as the starting point of our presentation 
in section~\ref{sec_act}.  A gauge invariant regularization is 
obtained by coupling two scalar fields \cite{refB} and leads to 
our model by freezing these fields up to their gauge transformations, 
here called London \cite{Lo35} limit. The scalar fields can then be
absorbed by extended gauge transformations, which were introduced
and discussed in \cite{Be09a,Be09b}. Some technical details 
of section~\ref{sec_act} encountered in the investigation of 
the classical continuum limit are relegated to appendixes. 

In section~\ref{sec_Num} we investigate the quantum field theory 
numerically in the lattice regularization. Evidence for a SU(2) 
deconfining phase transition, first observed in \cite{Be09a}, and 
results for the particle spectrum are presented. A massless photon 
is found in both phases. Noisy correlations (as usual) indicate 
massive SU(2) glueballs, which appear to jump to unphysically 
high mass values in the deconfined phase. A beautiful strong signal 
for correlation functions of a massive vector boson triplet is seen 
in the deconfined phase, while it does not exist on the confined side.

Summary, outlook and conclusions follow in the final 
section~\ref{sec_sum}.

\section{Action and Gauge Transformations \label{sec_act}}

Introducing gauge fields is often motivated by promoting a global 
symmetry of matter fields to a local one. Let us assume a global 
U(2) symmetry:
\begin{eqnarray} \label{G}
  \psi\ \to\ \psi{\,'} = G\,\psi &{\rm with}& G=G_1G_2\,,\\ 
  G_1\in U(1) &{\rm and}& G_2\in SU(2)\,. \nonumber
\end{eqnarray}
The $\overline{\psi} \gamma_{\mu}\partial_{\mu} \psi$ part of the free 
fermion Lagrangian breaks (\ref{G}) as a local symmetry. This is 
overcome by coupling to gauge fields, 
whose transformation behavior cancels the unwanted contributions out.
In the lattice regularization the relevant coupling reads
\begin{equation} \label{Upsi}
  a^3 K_{\mu}\,\overline{\psi}(na)\,U_{\mu}(na)
  V_{\mu}(na)\psi(na+\hat{\mu}a)\,,
\end{equation}
where $n=(n_1,n_2,n_3,n_4)$ is a vector of integers, $\hat{\mu}$ is 
the unit vector in $\mu$ direction, $a$ the lattice spacing ($x=na$),
\begin{equation} \label{UV}
  U_{\mu}=e^{ig_aaA_{\mu}}\in U(1)\ {\rm and}\
  V_{\mu}=e^{ig_baB_{\mu}}\in SU(2)\,.
\end{equation}
Here $g_a$ and $g_b$ are introduced as U(1) and SU(2) coupling 
constants. The precise definition of the fermions (see for 
instance \cite{MM94,Ro05}) in Eq.~(\ref{Upsi}) is accomplished by 
$K_{\mu}$ and irrelevant in our context as we discuss only their 
gauge transformations. The classical continuum limit is for 
$a\to 0$ obtained from $a^4$ contributions, whose sum becomes 
an integration over space.

In the lattice formulation it is obvious that the local gauge 
transformations
\begin{equation} \label{Gpsi}
  \psi(na)\ \to\ G(na)\,\psi(na) = G_1(na)\,G_2(na)\,\psi(na)
\end{equation}
will be absorbed by imposing the transformations
\begin{eqnarray} \label{GU}
  U_{\mu}(na) &\to& G_1(na) U_{\mu}(na) G_1^{-1}(na+\hat{\mu}a)\,,
  \\ \label{GV}
  V_{\mu}(na) &\to& G_2(na) V_{\mu}(na) G_2^{-1}(na+\hat{\mu}a)\,,
\end{eqnarray}
on the vector fields, provided that their contributions to the action 
are gauge invariant. This is (for instance) fulfilled for the Wilson 
action
\begin{equation} \label{Sgauge}
  S^{\rm gauge} = \frac{\beta_a}{2} \sum_p {\rm Re}\,{\rm Tr}\,U_p
    + \frac{\beta_b}{2} \sum_p {\rm Tr}\,V_p
\end{equation}
where the sums are over all plaquettes and $U_p$, $V_p$ are oriented 
products of gauge matrices around the plaquette loop. E.g., for a 
$V_p$ plaquette in the $\mu\nu$, $\mu\ne\nu$ plane:
$$V_{\mu\nu}(x) = {\rm Tr}\,\left[V_{\mu}(x)\,V_{\nu}(x+\hat{\mu}a)
     \,V^{\dagger}_{\mu}(x+\hat{\nu}a)\,V^{\dagger}_{\nu}(x)\right]\,.$$
As it turns out to be convenient, we represent here and in the 
following U(1) phase factors by diagonal $2\times 2$ matrices 
with $a_{\mu}$ (\ref{Sew}) and $A_{\mu}$ (\ref{UV}) related by 
\begin{equation} \label{Amu}
  A_{\mu} = \frac{1}{2}\,\tau_0\,a_{\mu}\,,
\end{equation}
where $\tau_0$ is the $2\times 2$ unit matrix. The electromagnetic
contribution to the Lagrangian (\ref{Sew}) reads then
\begin{eqnarray} \label{La}
  L^a~=~-\frac{1}{2}\,{\rm Tr}\,F^a_{\mu\nu}F^a_{\mu\nu}\,,
  ~~F^a_{\mu\nu}~=~\partial_{\mu}A_{\nu}-\partial_{\nu}A_{\mu}\,.
\end{eqnarray}
The equations 
\begin{equation} \label{babb}
  \beta_a=\frac{1}{g_a^2}~~{\rm and}~~\beta_b=\frac{4}{g_b^2}\,.
\end{equation}
relate $\beta_a$ and $\beta_b$ of (\ref{Sgauge}) to the couplings 
$g_a$ and $g_b$ of~(\ref{UV}). More lattice actions that give the 
classical continuum limit (\ref{Sew}) are discussed in~\cite{Ro95}.

There is an asymmetry in the way the gauge transformations work: 
A transformation of either the U(1) or the SU(2) matrices implies the 
corresponding transformation for the fermion field, while there is no 
such effect on the other gauge field. Another remarkable \cite{Sh86} 
feature of the action (\ref{Sgauge}) is that it is additive and not 
multiplicative in the U(1) and SU(2) fields. Changing also the fermion 
terms (\ref{Upsi}) to additive 
\begin{eqnarray} \label{addpsi}
  &~& a^3 K_{\mu}\,\overline{\psi}(na)\,U_{\mu}(na)
  \psi(na+\hat{\mu}a) \\
  &+& a^3 K_{\mu}\,\overline{\psi}(na)\,V_{\mu}(na)
  \psi(na+\hat{\mu}a)\,, 
\end{eqnarray}
the $a^4$ contributions to the classical $a\to 0$ limit remain 
unchanged, while the gauge transformations become symmetrical in all 
fields, because Eq.~(\ref{GU}) and (\ref{GV}) have to be replaced by
local U(2) transformations
\begin{eqnarray} \label{aGU}
  U_{\mu}(na) &\to& G(na) U_{\mu}(na) G^{-1}(na+\hat{\mu}a)\,,
  \\ \label{aGV}
  V_{\mu}(na) &\to& G(na) V_{\mu}(na) G^{-1}(na+\hat{\mu}a)\,,
\end{eqnarray}
which we call {\it extended gauge transformations} while reserving
the notation {\it gauge transformations} for their conventional version.

The gauge part (\ref{Sgauge}) of the action is invariant under local
U(2) transformations. After a generic transformation former U(1) and 
SU(2) matrices will both be in U(2) and their identification as U(1) 
or SU(2) has become converted into constraints on the gauge manifold. 
The special gauge in which they are actually in U(1) and SU(2), i.e. 
Eq.~(\ref{UV}) holds, will be called {\it diagonal gauge}. In this gauge
the U(1) part is not only diagonal but proportional to the unit matrix.
With reference to the existence of this gauge we keep the U(1) and SU(2) 
matrix identifications.

Requiring invariance only under extended gauge transformations 
(\ref{aGU}) and (\ref{aGV}) allows for new additions to the action. 
Numerically we investigate 
\begin{eqnarray} \label{Sint} 
  S^{\rm add}\ =\ \sum_{\mu\nu} S_{\mu\nu}^{\rm add}\,,~~~
  S^{\rm add}_{\mu\nu}\ = \qquad \qquad \qquad \qquad \qquad 
  \\ \label{Spint} \frac{\lambda}{2}\,{\rm Re}\,{\rm Tr}\,
  \left[U_{\mu}(x)\,V_{\nu}(x+\hat{\mu}a)\,U^{\dagger}_{\mu}
  (x+\hat{\nu}a)\,V^{\dagger}_{\nu}(x)\right]\,.~~~
\end{eqnarray}
A gauge invariant regularization is \cite{refB}
\begin{eqnarray} \label{Sscalar}
  S^s &=& \sum_x\left\{\sum_{\mu\nu}S^s_{\mu\nu}+\kappa\,{\rm Tr}\,
  \left[\left(\Phi^{\dagger}\,\Phi-\tau_0\right)^2\right]\right\},~~~~~
  \\ \label{Ssp} S_{\mu\nu}^s& = & \frac{\lambda^s}{4} {\rm Re}\,
  {\rm Tr}\,\{U_{\mu}(x) U^{\dagger}_{\mu}(x+\hat{\nu}a)\\ \nonumber && 
     [\Phi^{\dagger}(x+\hat{\mu}a) 
  \,V_{\nu}(x+\hat{\mu}a)\,\Phi(x+\hat{\mu}a+ \hat{\nu}a)]\\ \nonumber 
  && [\Phi^{\dagger}(x)\,V_{\nu}(x)\,\Phi(x+\hat{\nu}a)]^{\dagger}\}
\nonumber
\end{eqnarray}
where $\Phi$ is a $2\times 2$ matrix scalar field that is charged 
with respect to $U(1)$ and  $SU(2)$. Its gauge transformations are 
\begin{equation} \label{PhiGauge} 
  \Phi\ \rightarrow\ e^{-i\alpha}\, g\, \Phi\,, 
\end{equation}
where $g\in\rm SU(2)$, $e^{i\alpha}\in\rm U(1)$. In the London 
\cite{Lo35} limit $\kappa \to\infty$ this action is equivalent 
to (\ref{Sint}), because $\Phi$ takes on it vacuum value, which 
is a pure gauge 
\begin{equation} \label{pureGauge} 
  \Phi\ =\ e^{-i\alpha}\,g\,. 
\end{equation}
Fixing the gauge to $\Phi=\tau_0$ yields~(\ref{Sint}) and on finite
lattices properties at sufficiently large $\kappa$ values are expected 
to be practically identical.

It is only in the limit $\kappa \to\infty$ that the coupling constant 
$\lambda^s$ together with the dimensions of the scalar fields becomes 
the dimensionless coupling $\lambda$ of~(\ref{Sint}) (recall that the 
dimensionless lattice field $\Phi$ is related to the physical scalar 
field $\Phi_{\rm phys}$ by $\Phi=a\,\Phi_{\rm phys}$ \cite{Ro05}). 
This is opposite to what happens in the SU(2) Higgs model where the 
interaction between scalar field and SU(2) field, as for instance given 
in \cite{MMH}, is proportional to
\begin{equation} \label{SU2Higgs} 
  \lambda^s_h \sum_x \sum_{\mu=1}^4{\rm Tr}\,\left(\phi^{\dagger}_{x
  +\hat{\mu}a}\,V_{\mu}(x)\,\phi_x\right)\,,
\end{equation}
with $\phi=\rho\,g$, $\rho>0$, $g\in$ SU(2). Here $\lambda^s_h$ is
dimensionless, and fixing the gauge in the London limit to $\phi=\tau_0$ 
leads to the gauge symmetry breaking contribution 
\begin{equation} \label{SU2broken} 
  \lambda_h\sum_x \sum_{\mu=1}^4 {\rm Tr}\, V_{\mu}(x)\,, 
\end{equation}
where $\lambda_h$ acquires the dimension of the $\phi^{\dagger}\phi$ 
scalar fields.

The extended gauge transformations (\ref{aGU}) and (\ref{aGV}) restore
the transformation (\ref{PhiGauge}) without using scalar fields by 
absorbing the $\exp(-i\alpha)$ phase factor into the $V_{\mu}$ and 
the SU(2) transformation $g$ into the $U_{\mu}$ field. This requires 
the action (\ref{Sint}), or similar, and is not possible in the London 
limit (\ref{SU2broken}) of the Higgs model.

Before investigating the quantum field theory (\ref{Sint}) numerically, 
we consider the classical continuum limits of (\ref{Sint}) 
and~(\ref{Sscalar}).

\subsection{Classical continuum limit}

The classical Lagrangian is expected to be the effective 
theory at energy scales much larger than the masses of the model.
In the limit $a\to 0$ the $a^4$ contributions of (\ref{Spint}) give
\begin{eqnarray} \label{Lint} 
  L^{\rm add} &=& -\frac{\lambda}{4}\,{\rm Tr}\left(
  F^{\rm add}_{\mu\nu}F^{\rm add}_{\mu\nu}\right)\,,\\ \label{Fint}
  F^{\rm add}_{\mu\nu} &=& g_b\partial_{\mu}B_{\nu} -
  g_a\partial_{\nu}A_{\mu} + i\,g_ag_b\left[A_{\mu},B_{\nu}\right]\,,
\end{eqnarray}
where the commutator reflects that $B_{\mu}$ and $A_{\nu}$ will in
general not commutate unless we are in the diagonal gauge, defined 
in-between Eq.~(\ref{aGV}) and~(\ref{Sint}). 

Let us translate the extended gauge transformations (\ref{aGU}) and 
(\ref{aGV}) into continuum notation. Consider the gauge-covariant 
derivative 
\begin{equation} \label{Dem} 
  D^{\rm em}_{\mu} = \partial_{\mu} + ig_a a_{\mu}
\end{equation}
of an electromagnetic field $a_{\mu}(x)$ on a complex fermion field 
$\psi(x)$. With the gauge transformations
\begin{eqnarray} 
  \psi &\to& \psi' = e^{i\alpha(x)}\,\psi\,, \\
  a_{\mu}(x)&\to& a'_{\mu} =
  a_{\mu}-\frac{i}{g_a}\,\partial_{\mu}\alpha(x)
\end{eqnarray}
one finds
\begin{equation}
  D^{\rm em}_{\mu}\psi\to D^{'\rm em}_{\mu}\psi'
  = e^{i\alpha(x)}D^{\rm em}_{\mu}\psi\,,
\end{equation}
so that the Lagrangian
\begin{equation} 
  L^{\rm em}_{\psi} = 
  \overline{\psi}\left(i\gamma_{\mu}D^{em}_{\mu}-m\right)\psi 
  - \frac{1}{4}F^{\rm em}_{\mu\nu}F^{\rm em}_{\mu\nu}
\end{equation}
is gauge invariant, where 
\begin{equation} \label{Fem} 
  F^{\rm em}_{\mu\nu}=\frac{1}{ig_a}
  \left[D^{em}_{\mu},D^{em}_{\nu}\right]\,.
\end{equation}
is the field tensor.

Assume now that $\psi(x)$ is a complex doublet, which transforms
under local U(2) transformations
\begin{eqnarray} \label{psig}
  \psi\ &\to& \psi' = G(x)\,\psi\,,\\ \label{Gcont}
  G(x) &=& \exp\left(\frac{i}{2}
  \sum_{i=0}^3\tau_i\alpha_i(x)\right)\,.
\end{eqnarray}
Here $\tau_0$ is as in (\ref{Amu}) the $2\times 2$ unit matrix and 
$\tau_i$, $i=1,2,3$ are the Pauli matrices. We still can couple $\psi$ 
in an invariant way to an electromagnetic field. We define the gauge 
covariant derivative by
\begin{equation} \label{Da} 
  D^a_{\mu} = \partial_{\mu} + ig_a A_{\mu}
\end{equation}
and the extended gauge transformations of $A_{\mu}$ by (compare for
instance Ref.~\cite{Qu83} for usual SU(2) gauge transformations)
\begin{equation} \label{Ag} 
  A_{\mu}\to A'_{\mu} = GA_{\mu}G^{-1}
  + \frac{i}{g_a}(\partial_{\mu} G)\,G^{-1}
\end{equation}
which adds to the U(1) field the transformation of a null SU(2) field 
and yields the desired result
\begin{equation} \label{Dpsig}
  D^a_{\mu}\psi\to D^{'a}_{\mu}\psi' = G D^a_{\mu}\psi\,.
\end{equation}
The field tensor defined by
\begin{equation} 
  F^a_{\mu\nu}=\frac{1}{ig_a}\left[D^a_{\mu},D^a_{\nu}\right]
\end{equation}
transforms as
\begin{equation} \label{Fag} 
  F^a_{\mu\nu}\to F^{'a}_{\mu\nu} = G F^a_{\mu\nu} G^{-1}
\end{equation}
so that the Lagrangian 
\begin{equation} \label{Lapsi} 
  L^a_{\psi} = \overline{\psi}\left(i\gamma_{\mu}D^a_{\mu}-m\right)\psi 
      -\frac{1}{2}{\rm Tr}\left(F^a_{\mu\nu}F^a_{\mu\nu}\right)
\end{equation}
stays invariant. The local U(2) transformations do
not destroy the fact that $A_{\mu}$ describes just an electromagnetic 
field. Any $A_{\mu}(x)$ field is gauge equivalent to one in the 
diagonal gauge for which $A_{\mu}$ is proportional to the unit matrix. 
Consequently, $F^{'a}_{\mu\nu}$ stays always diagonal and the $G$ 
matrices can be omitted in~(\ref{Fag}). 

Similarly, we extend gauge transformations of a SU(2) field
$B_{\mu}(x)$ by a phase. The gauge covariant derivative is
\begin{equation} \label{Db} 
  D^b_{\mu} = \partial_{\mu} + ig_b B_{\mu}~~{\rm with}~~
  B_{\mu} = \frac{1}{2}\,\vec{\tau}\cdot\vec{b}_{\mu}
\end{equation}
and the extended gauge transformations of $B_{\mu}$ are
\begin{equation} \label{Bg} 
  B_{\mu}\to B'_{\mu} = GB_{\mu}G^{-1}
  + \frac{i}{g_b}(\partial_{\mu} G)\,G^{-1}\,.
\end{equation}
Equations (\ref{Dpsig}) to (\ref{Lapsi}) carry simply over by
replacing all labels $a$ by~$b$.

An electroweak Lagrangian of the type
\begin{eqnarray} \label{Labpsi}
  L^{ab}_{\psi}&=&-\frac{1}{2}{\rm Tr}\left(F^a_{\mu\nu}F^a_{\mu\nu}
  \right) -\frac{1}{2}{\rm Tr}\left(F^b_{\mu\nu}F^b_{\mu\nu}\right)
  \\ \nonumber
  &+& \overline{\psi}\left(i\gamma_{\mu}D^a_{\mu}-m\right)\psi 
   +  \overline{\psi}\left(i\gamma_{\mu}D^b_{\mu}-m\right)\psi
\end{eqnarray}
allows one to add (\ref{Lint}). Under extended gauge transformations 
(\ref{Ag}) for $A_{\mu}$ and (\ref{Bg}) for $B_{\mu}$, the 
$F^{\rm add}_{\mu\nu}$ tensor (\ref{Fint}) transforms according to 
\begin{equation} \label{Fintg} 
  F^{\rm add}_{\mu\nu}\to F^{'\rm add}_{\mu\nu} = 
  G\, F^{\rm add}_{\mu\nu}\, G^{-1}\,,
\end{equation}
so that $L^{\rm add}$ is invariant. The algebra for (\ref{Fintg}) 
is given in appendix~\ref{appA}.

As it is tedious and not very enlightening to calculate the classical 
continuum limit of the action (\ref{Sscalar}) with scalar fields, we 
relegate its discussion to appendix~\ref{appB}. The essence is caught 
by a much simpler gauge invariant classical interaction with two scalar 
fields, which reduces also to (\ref{Lint}) in the London limit, but has 
a lattice regularization that differs from~(\ref{Sscalar}).

Let us introduce the scalar fields
\begin{equation} \label{phi1phi2} 
  \phi_1 = \rho\,e^{i\alpha}\,\tau_0,\ \rho>0~~{\rm and}~~
  \phi_2 = \sigma\,g,\ \sigma>0
\end{equation}
with $g\in\rm SU(2)$. It is then easy algebra to show that the field 
tensor
\begin{eqnarray} \nonumber 
  S^s_{\mu\nu}&=&\phi_2^{\dagger}\,D^b_{\mu}\,D^b_{\nu}\,\phi_2
   + (D^b_{\mu}\,\phi_2)^{\dagger}\,D^b_{\nu}\,\phi_2 \\ \label{Suv} 
  &-&\phi_1^{\dagger}\,D^a_{\nu}\,D^a_{\mu}\,\phi_1
   - (D^a_{\nu}\,\phi_1)^{\dagger}\,D^a_{\mu}\,\phi_1
\end{eqnarray}
reduces by gauge fixing in the London limit to $i\,F^{\rm add}_{\mu\nu}$ 
(\ref{Fint}) in the diagonal gauge, so that
\begin{equation} \label{LSint} 
  L^s = - \frac{\lambda}{4}\,{\rm Tr}\,
  \left[\left(S^s_{\mu\nu}\right)^{\dagger}\,S^s_{\mu\nu}\right]
\end{equation}
becomes a gauge invariant extension of (\ref{Lint}).

\section{Numerical investigation \label{sec_Num}}

Pure U(1) LGT with the Wilson action has at $\beta_a\approx 1.01$ 
a phase transition from its confined into its Coulomb phase. Presumably
the transition is weakly first order as first reported in \cite{JeNe83}. 
We have calculated lattice averages of U(1) Polyakov loops $P_a$ by 
Monte Carlo (MC) simulations on a $12^4$ lattice with periodic boundary 
conditions with a statistics of 10,000 sweeps for equilibration and 
160,000 sweeps with measurements. Measurements are plotted every 20 
sweeps. Figure~\ref{fig_U1scatter1} compares scatter plots at $\beta_a
=0.9$ in the disordered confined phase and at $\beta_a=1.1$ in the 
ordered Coulomb phase. In the unbroken disordered phase the values 
scatter about zero, while they form a ring in the broken ordered 
phase.  As the transition happens at zero temperature, this holds 
on a symmetric lattice for Polyakov loops winding in any one of 
the four directions through the torus (ordered starts are used 
to avoid metastabilities of the MC algorithm). 

In the following MC simulations with the plaquette action
\begin{equation} \label{Smodel} 
  S\ =\ S^{\rm gauge} + S^{\rm add}
\end{equation}
are performed, where $S^{\rm gauge}$ is defined by (\ref{Sgauge}) 
and $S^{\rm add}$ by (\ref{Sint}). We use lattice units $a=1$ in 
this section.

\subsection{Integration measure and Monte Carlo updating \label{sec_MC}}

Our MC procedure proposes the usual U(1) and SU(2) changes. For the
update of a U(1) matrix $U_{\mu}(n)$ we need the contribution to 
(\ref{Smodel}), which comes from the eight staples containing this matrix
\begin{eqnarray} \nonumber 
  U_{\sqcup,\mu}(n) &=& \frac{\beta_a}{2}\sum_{\nu\ne\mu} 
  \left[U_{\nu}(n+\hat{\mu})\,U^{\dagger}_{\mu}(n+\hat{\nu})\,
  U^{\dagger}_{\nu}(n)\,\right. \\ \nonumber
  &+& \left. U^{\dagger}_{\nu}(n+\hat{\mu}-\hat{\nu})\,
  U^{\dagger}_{\mu}(n-\hat{\nu})\, U_{\nu}(n-\hat{\nu})\right]\\ 
  \nonumber &+& \frac{\lambda}{2}\sum_{\nu} \left[ 
  V_{\nu}(n+\hat{\mu})\,U^{\dagger}_{\mu}(n+\hat{\nu})\,
  V^{\dagger}_{\nu}(a)\,\right. \\ \label{U1staples}
  &+& \left. V^{\dagger}_{\nu}(n+\hat{\mu}-\hat{\nu})\,
  U^{\dagger}_{\mu}(n-\hat{\nu})\, V_{\nu}(n-\hat{\nu})\right]~~~~~
\end{eqnarray}
and correspondingly for the SU(2) matrix $V_{\mu}(n)$
\begin{eqnarray} \nonumber
  V_{\sqcup,\mu}(n) &=& 
  \frac{\beta_b}{2}\sum_{\nu\ne\mu} \left[ 
  V_{\nu}(n+\hat{\mu})\,V^{\dagger}_{\mu}(n+\hat{\nu})\,
  V^{\dagger}_{\nu}(n)\,\right. \\ \nonumber
  &+& \left. V^{\dagger}_{\nu}(n+\hat{\mu}-\hat{\nu})\,
  V^{\dagger}_{\mu}(n-\hat{\nu})\, V_{\nu}(n-\hat{\nu})\right]\\ 
  \nonumber &+& \frac{\lambda}{2}\sum_{\nu} \left[ 
  U_{\nu}(n+\hat{\mu})\,V^{\dagger}_{\mu}(n+\hat{\nu})\,
  U^{\dagger}_{\nu}(n)\,\right. \\ \label{SU2staples}
  &+& \left. U^{\dagger}_{\nu}(n+\hat{\mu}-\hat{\nu})\,
  V^{\dagger}_{\mu}(n-\hat{\nu})\, U_{\nu}(n-\hat{\nu})\right]\,.~~~~~
\end{eqnarray}
Updates are then performed with the Gibbs-Boltzmann weights
\begin{eqnarray} \label{U1w}
  &dU_{\mu}(n)&\exp\left\{\,{\rm Re}{\rm Tr}\left[U_{\mu}(n)\,
  U_{\sqcup,\mu}(n) \right]\,\right\}\,,\\ \label{SU2w}
  &dV_{\mu}(n)&\exp\left\{\,{\rm Re}{\rm Tr}\left[V_{\mu}(n)\,
  V_{\sqcup,\mu}(n) \right]\,\right\}\,.
\end{eqnarray}
where the $dU_{\mu}(n)$ integration is from $-\pi$ to $+\pi$ over the 
phase $\phi_{\mu}(n)$ of the matrix $U_{\mu}(n)=\exp[i\,\phi_{\mu}(n)]$ 
and $dV_{\mu}(n)$ is over the SU(2) Hurwitz \cite{Hu97} measure. This
is well suited and done with the biased Metropolis-heatbath 
algorithm~\cite{Ba05}. 

It is instructive to discuss (\ref{SU2w}) for the case in which the 
U(1) matrices in (\ref{SU2staples}) are aligned as it is approximately
the case in the U(1) Coulomb phase. Then the U(1) matrices cancel out, 
so that large values of $\lambda$  favor ${\rm Tr} [V_{\mu}(n)
V^{\dagger}_{\mu}(n+\hat {\nu})]={\rm Tr} [V_{\mu}(n) 
V^{\dagger}_{\mu}(n-\hat{\nu})]=2$ and, hence, the SU(2) 
matrices are also aligned, $V_{\mu}(n)=
V_{\mu}(n+\hat{\nu})=V_{\mu}(n-\hat{\nu})$.  SU(2) deconfinement 
(breaking of the Z$_2$ center group) is achieved when this effect 
(spoiled by U(1) fluctuations at finite $\lambda$) becomes strong 
enough. Integrations with the measures (\ref{U1w}) and (\ref{SU2w}) 
do not include extended gauge transformations (\ref{aGU}) and 
(\ref{aGV}). Updates stay within the diagonal gauge defined in 
section~\ref{sec_act}. Within the MC procedure only global SU(2) 
transformations $V_{\mu}(n)\to GV_{\mu} (n)G^{-1}$ with the same 
$G$ for all SU(2) matrices remain.

The partition function of the MC calculation
\begin{equation} \label{Zmodel} 
  Z\{G(n)\}\ =\ \int \prod_n\prod_{\mu=1}^4 dU_{\mu}(n)\,dV_{\mu}(n)
  \, e^{S\{U_{\mu}(n),V_{\mu}(n)\}}
\end{equation}
is invariant under extended $U(2)$ gauge transformation (\ref{aGU}) 
and (\ref{aGV}). The Jacobian determinants of both $\prod_n\prod_{\mu} 
dU_{\mu}(n)$ and $\prod_n\prod_{\mu} dV_{\mu}(n)$ are one and the 
action (\ref{Smodel}) is by construction invariant. U(1) and SU(2) 
Wilson loops are invariant operators as the U(2) transformations 
drop out in the trace. As usual \cite{MM94,Ro05} their 
physical interpretation can be derived by imagining a coupling to 
static quarks. In addition to the normal Wilson loops traces of 
corresponding mixed products of U(1) and SU(2) matrices are also 
invariant, but we make no use of them in the following.

With normalization of the integration over the $U(2)$ measure to 
one, the identity
\begin{equation} \label{Gint} 
  Z\ =\ \int \prod_n dG(n)\,Z\{G(n)\}\ =\ Z\{G(n)\}
\end{equation}
holds under extended gauge transformations. This integration can 
easily be added to the updates: At site $n$ all matrices on links 
emerging at $n$ will be transformed according to
\begin{equation} \label{GUGV} 
  U_{\mu}(n)\to G(n)\,U_{\mu}(n)\,,~~~
  V_{\mu}(n)\to G(n)\,V_{\mu}(n)
\end{equation}
and all matrices on links ending at $n$ according to
\begin{eqnarray} \nonumber
  U_{\mu}(n-\hat{\mu})&\to& U_{\mu}(n-\hat{\mu})\,G^{-1}(n)\,,
  \\ \label{UGVG}
  V_{\mu}(n-\hat{\mu})&\to& V_{\mu}(n-\hat{\mu})\,G^{-1}(n)\,.
\end{eqnarray}
The acceptance rate of such updates is 100\% as no change in the 
value of the action is implied.

The unusual feature is that the integration measure of the vector
fields $U_{\mu}(n)$ and $V_{\nu}(n)$ in the functional integral 
(\ref{Zmodel}) is not identical with the gauge measures in 
(\ref{Gint}). However, for gauge transformations of scalar and 
fermion fields we are used to this and there appears no strong
argument against having a distinction of these measures also for
vector fields (calling them gauge fields may have added to the 
expectation that functional and gauge measures are the same).

\subsection{Zero temperature SU(2) deconfining transition 
\label{sec_deconf}}

\begin{figure}[tb] \begin{center} 
\epsfig{figure=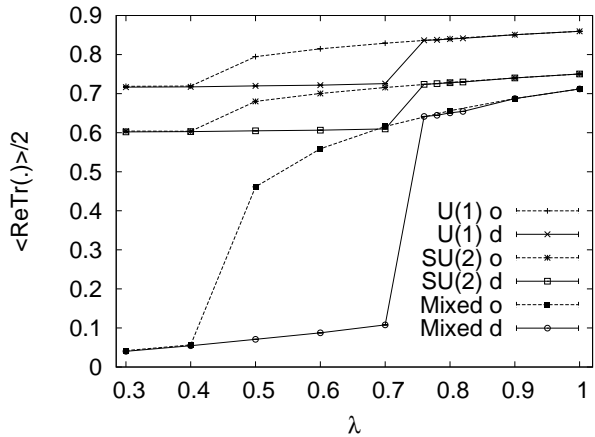,width=\columnwidth} 
\caption{Plaquette expectation values on a $12^4$ lattice as function 
of $\lambda$ with $\beta_a=1.1$ and $\beta_b=2.3$ (ordered o and
disordered d starts).  \vspace{-4mm} \label{fig_action} } \end{center} 
\end{figure}

\begin{table}[tb]
\caption{Plaquette expectation values of Fig.~\ref{fig_action}.
\label{tab_action}} 
\centering
\begin{tabular}{|c|c|c|c|c|} \hline
$\lambda$ & U(1) & SU(2) & Mixed   \\ \hline
0.30 d& 0.716576 (17)& 0.601961 (29)& 0.040142 (08)  \\ \hline
0.40 d& 0.717373 (17)& 0.602630 (27)& 0.054590 (08)  \\ \hline
0.50 d& 0.719702 (21)& 0.604709 (21)& 0.070845 (08)  \\ \hline
0.60 d& 0.721775 (19)& 0.606398 (22)& 0.087592 (10)  \\ \hline
0.70 d& 0.725563 (21)& 0.609602 (29)& 0.107814 (21)  \\ \hline
0.76 d& 0.836426 (08)& 0.723714 (08)& 0.641590 (14)  \\ \hline
0.78 d& 0.837722 (08)& 0.725532 (09)& 0.644591 (14)  \\ \hline
0.80 d& 0.839750 (07)& 0.727838 (09)& 0.650849 (10)  \\ \hline
0.82 d& 0.841460 (34)& 0.729742 (11)& 0.654499 (56)  \\ \hline
0.90 d& 0.850887 (07)& 0.740039 (07)& 0.687311 (11)  \\ \hline
1.00 d& 0.859572 (07)& 0.750205 (08)& 0.712059 (07)  \\ \hline
0.30 o& 0.718566 (19)& 0.603975 (21)& 0.042148 (06)  \\ \hline
0.40 o& 0.719359 (19)& 0.603618 (25)& 0.056604 (09)  \\ \hline
0.50 o& 0.794571 (11)& 0.680028 (10)& 0.461244 (39)  \\ \hline
0.60 o& 0.814685 (11)& 0.700420 (07)& 0.558826 (22)  \\ \hline
0.70 o& 0.829112 (08)& 0.715794 (08)& 0.615873 (14)  \\ \hline
0.80 o& 0.840897 (09)& 0.728681 (09)& 0.656371 (15)  \\ \hline
0.90 o& 0.850874 (06)& 0.740036 (08)& 0.687314 (07)  \\ \hline
1.00 o& 0.859553 (08)& 0.750204 (08)& 0.712054 (09)  \\ \hline
\end{tabular} \end{table} 
 
We keep the U(1) coupling at $\beta_a=1.1$ and for the SU(2) coupling 
the values $\beta_b=2.3$ is used. At $\lambda=0$, without interaction, 
$\beta_a$ is in the U(1) Coulomb phase and $\beta_b$ in the SU(2) 
scaling region. Runs are performed on $12^4$ lattices. Data points 
are based on a statistics of at least $2^{10}$ sweeps without 
measurements and subsequently $32\times 2^{10}$ sweeps with 
measurements. Error bars are calculated with respect to the 32 blocks.

With increasing $\lambda$ one finds a strong first order phase 
transition, which is illustrated in Fig.~\ref{fig_action} for the 
expectation values of the various plaquette actions. From up to 
down: $\langle{\rm Re Tr}U_p\rangle/2$ for U(1), $\langle{\rm Tr}V_p
\rangle/2$ for SU(2) both contributing to (\ref{Sgauge}) and 
$\lambda^{-1}\sum_{\mu\nu}\langle S_{\mu\nu}^{\rm add}\rangle/16$ 
for Mixed (\ref{Sint}), which drives the transition. 
Disordered starts are marked by ``d'' and ordered starts 
by ``o''. The values are compiled in table~\ref{tab_action}.

\begin{figure}[tb] \begin{center} 
\epsfig{figure=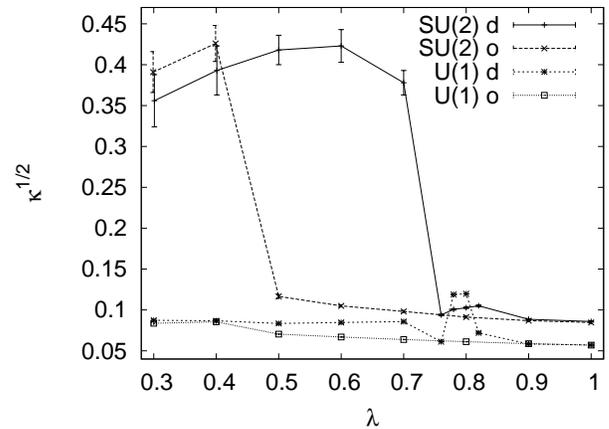,width=\columnwidth} 
\caption{SU(2) and U(1) string tensions from Creutz ratios on a $12^4$ 
lattice as function of $\lambda$ with $\beta_a=1.1$ and $\beta_b=2.3$
(disordered d and ordered o starts). \vspace{-4mm} \label{fig_k} } 
\end{center} \end{figure}

\begin{table}[tb]
\caption{String tension values $\sqrt{\kappa}$ of Fig.~\ref{fig_k}.
\label{tab_k}} 
\centering
\begin{tabular}{|c|c|c|c|c|c|} \hline
$\lambda$& U(1) d & SU(2) d    & U(1) o      & SU(2) o    \\ \hline
0.30& 0.0873  (16)& 0.356  (32)& 0.0837  (13)& 0.391  (25)\\ \hline
0.40& 0.0868  (18)& 0.393  (30)& 0.0855  (16)& 0.426  (22)\\ \hline
0.50& 0.0835  (16)& 0.418  (18)& 0.07045 (64)& 0.1166 (25)\\ \hline
0.60& 0.0847  (17)& 0.423  (20)& 0.06694 (45)& 0.1050 (13)\\ \hline
0.70& 0.0858  (16)& 0.378  (15)& 0.06388 (40)& 0.0982 (13)\\ \hline
0.76& 0.06117 (35)& 0.0940 (10)& $-$         & $-$        \\ \hline
0.78& 0.11887 (21)& 0.1007 (09)& $-$         & $-$        \\ \hline
0.80& 0.11979 (21)& 0.1027 (10)& 0.06124 (33)& 0.0913 (12)\\ \hline
0.82& 0.0720  (12)& 0.1051 (10)& $-$         & $-$        \\ \hline
0.90& 0.05839 (32)& 0.0886 (08)& 0.05874 (33)& 0.0869 (07)\\ \hline
1.00& 0.05714 (28)& 0.0860 (08)& 0.05704 (27)& 0.0847 (09)\\ \hline
\end{tabular} \end{table} 

Creutz ratios \cite{Cr80} for the SU(2) and U(1) string tensions are 
calculated from Wilson loops up to size $5\times 5$. Fig.~\ref{fig_k} 
shows for $\beta_b=2.3$ the behavior of the square roots of the string 
tensions in lattice units $a=1$ as function of $\lambda$. Error bars 
are calculated with respect to 32 jackknife bins, following the 
scheme of \cite{Be04}. While the jump in the plaquette actions is 
similar for U(1) and SU(2), this is not the case for the string 
tensions. For SU(2) $\sqrt{\kappa}$ decreases at the transition by a 
factor 3.5, whereas the drop of $\sqrt{\kappa}$ for U(1) is just 25\%.
The $\sqrt{\kappa}$ values are compiled in table~\ref{tab_k}.
 
\begin{figure}[tb] \begin{center}
\epsfig{figure=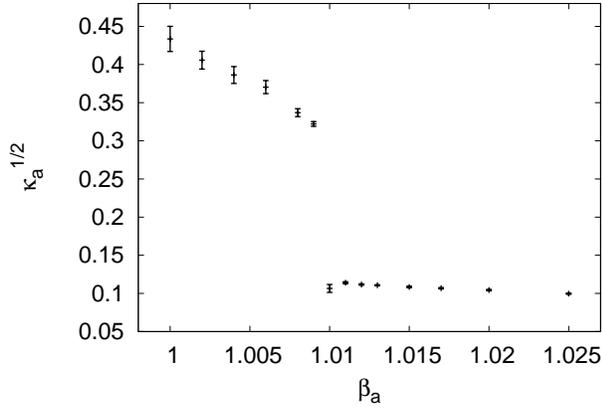,width=\columnwidth} 
\caption{U(1) string tension from Creutz ratios on a $12^4$ lattice
as function of $\beta_a$ ($\beta_b=\lambda=0$). 
\vspace{-4mm}\label{fig_U1k0} }\end{center}\end{figure}
 
The interpretation of Fig.\ref{fig_k} is that the U(1) string tension 
signals the deconfined phase on both sides of the transition, while 
the SU(2) string tension is characteristic for the confined phase at
small $\lambda$ and for a zero-temperature deconfined phase at large 
$\lambda$. The latter point is supported by comparison with the 
behavior of the U(1) string tension for $\beta_b=\lambda=0$ at the 
U(1) deconfining phase transition as shown in Fig.~\ref{fig_U1k0}. 
The discontinuity in the string tension is as in Fig.~\ref{fig_k} 
for SU(2), only that no strong metastabilities are observed for U(1).

\begin{figure}[tb] \begin{center}
\epsfig{figure=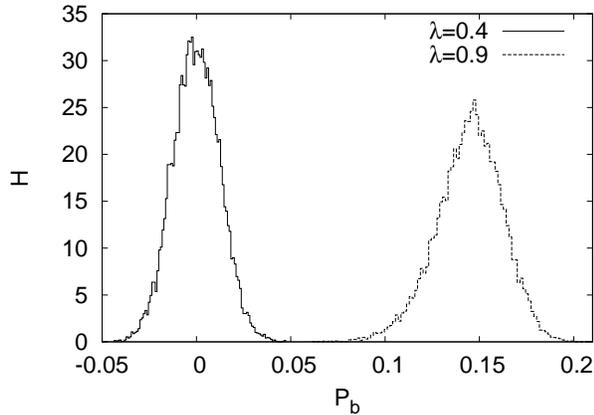,width=\columnwidth} 
\caption{Normalized SU(2) Polyakov loop histogram $H$ at $\beta_a=1.1$,
$\beta_b=2.3$ with $\lambda=0.4$ (left) and $\lambda=0.9$ (right). 
\vspace{-4mm} \label{fig_SU2Pol} } \end{center} \end{figure}

Polyakov loop measurements support also the SU(2) deconfining phase 
transition. Fig.~\ref{fig_SU2Pol} shows histograms for SU(2) Polyakov 
loops $P_b$ at $\lambda=0.4$ in the confined and at $\lambda=0.9$ 
in the deconfined phase. In the confined phase the values scatter 
symmetrically about zero, whereas in the deconfined phase the Z$_2$ 
center symmetry is broken and the values scatter about a mean of 
0.14458 (45). A very long run would produce a double peak at 
$\lambda=0.9$, but our run time was far too short to overcome the 
free energy barrier between the two peaks. Scatter plots of the U(1) 
Polyakov loops at the same couplings give in good approximation the 
ring of Fig.~\ref{fig_U1scatter1} at $\lambda=0.4$ and a more 
pronounced ring at $\lambda=0.9$, both deconfined. This interpretation
is confirmed by evidence for a massless photon on both sides and of 
interest because compact U(1) LGT allows in contrast to non-compact 
U(1) LGT also for a confined phase.

\subsection{Spectrum calculations \label{sec_spectrum}}

Mass spectrum calculations were performed on lattices of size $N^3 N_t$, 
$N_t>N$ in the range $4^3 16$ to $12^3 64$ with a statistics of $n_{\rm 
eq}$ sweeps for reaching equilibrium and $32\times n_{\rm eq}$ sweeps 
with measurements. Data analysis with respect to 32 bins follows again 
the jackknife scheme of~\cite{Be04}. The $n_{\rm eq}$ values used for 
different lattice sizes are contained in table~\ref{tab_photon}. All 
simulations reported in this section are at $\beta_a=1.1$ and $\beta_b
=2.3$.

\begin{table}[tb]
\caption{Lattice sizes, statistics and photon mass estimates 
at $\lambda=\lambda_1=0.4$ and $\lambda=\lambda_2=0.9$. 
\label{tab_photon}} 
\centering
\begin{tabular}{|c|c|c|c|c|} \hline
Lattice & $n_{\rm es}$ & $m^2_{\rm photon}(\lambda_1)$ &
          $m^2_{\rm photon}(\lambda_2)$ &$4\sin^2(k_1/2)$   \\ \hline
$4^3 16$& $2^{11}$& $-$0.252  (20)& $-$0.289  (21)& 2       \\ \hline
$6^3 24$& $2^{13}$& $-$0.080  (11)& $-$0.0728 (78)& 1       \\ \hline
$8^3 32$& $2^{14}$& $-$0.0138 (18)& $-$0.0248 (14)& 0.585786\\ \hline
$10^348$& $2^{14}$& $-$0.0100 (20)& $-$0.0123 (11)& 0.381966\\ \hline
$12^364$& $2^{14}$& $-$0.0034 (12)& $-$0.0061 (07)& 0.267949\\ \hline
\end{tabular} \end{table} 

\begin{figure}[tb] \begin{center}
\epsfig{figure=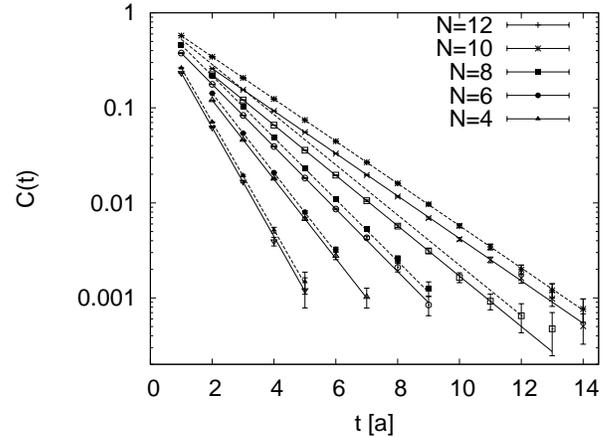,width=\columnwidth} 
\caption{Correlation functions data and fits for the photon mass 
estimates. The up-down order of the curves agrees with that of 
the labeling. The upper in a pair of curves is always at $\lambda=0.9$
and the lower at $\lambda=0.4$. \vspace{-4mm} \label{fig_Ek1} } 
\end{center} \end{figure}

First, we estimate the photon mass at $\lambda=0.4$ in the disordered 
SU(2) phase and in the ordered SU(2) phase at $\lambda=0.9$. Following 
\cite{BP84} this is done via the lattice dispersion relation (derived 
for a free field, for instance, in \cite{Ro05})
\begin{equation} \label{dispersion}
  m^2_{\rm photon}\ =\ E^2_{k_1} - 4\,\sin^2\left(\frac{k_1}{2}\right)
\end{equation}
where $E_{k_1}$ is the energy of the U(1) momentum eigenstate 
$(k_1,0,0)$, $k_1= 2\pi/N$, of the plaquette trial operator in the 
$T_1^{+-}$ representation of the cubic group. Energies are estimated 
from correlation functions $c(t)$ of operators by the usual two 
parameter ($a_1$ and $E$) $\cosh$ fits
\begin{equation} \label{coshE}
  c(t)\ =\ a_1\,\left(e^{-E\,t}+e^{-E\,(N_t-t)}\right)
\end{equation}
in a range $t_1\le t\le t_2$, so that the goodness of fit in the sense
discussed for correlated data in \cite{Be04} is acceptable. Our 
$E_{k_1}$ correlations functions at $\lambda=0.4$ and $\lambda=0.9$
are shown in Fig.~\ref{fig_Ek1}. The corresponding $m^2_{\rm photon}$
estimates are compiled in table~\ref{tab_photon}. The last column of 
the table gives the lowest non-zero lattice momentum squared.

While in \cite{BP84} the photon mass estimates from simulations on
$4^3 16$ lattices were within statistical errors consistent with zero, 
this is due to increased statistical accuracy no longer true, instead 
$m^2_{\rm photon}$ comes out negative. This holds not just for our 
action, but also for pure U(1) LGT, where we obtained $m^2_{\rm photon} 
= - 0.2102\ (17)$ on a $4^3 16$ lattice at $\beta_a=1.1$. Apparently 
we are dealing with an infrared cutoff effect, which disappears with 
increasing lattice size and becomes thus consistent with \cite{Ma04} 
where $m^2_{\rm photon}=0$ was again found within statistical errors
for pure U(1) LGT. So, we conclude that $m^2_{\rm photon} \to 0$ for 
$N\to\infty$ holds also in our cases.

\begin{figure}[tb] \begin{center}
\epsfig{figure=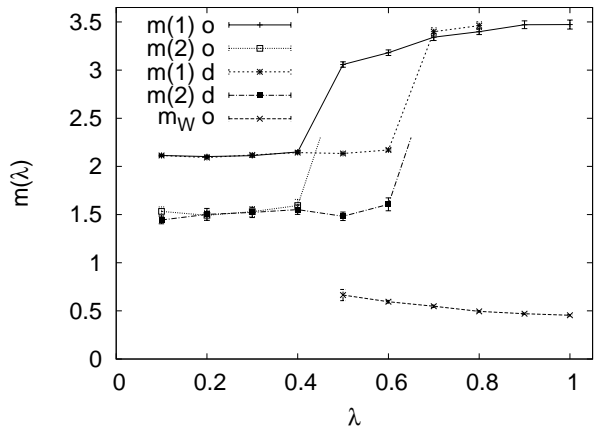,width=\columnwidth} 
\caption{Glueball and vector boson mass estimates on a $4^316$
lattice (o ordered and d disordered SU(2) starts). \vspace{-4mm} 
\label{fig_masses} } 
\end{center} \end{figure} 

For an overview of the SU(2) glueball spectrum a $4^3 16$ lattice 
appears to be large enough, because glueball mass values turn out 
to be high. Compared to the $E_{k_1}$ correlations, the signal from 
glueball correlation functions is very noisy, strongest for the lowest 
lying zero momentum $A_1^{++}$ representation of the cubic group. In 
Fig.~\ref{fig_masses} we give for the $A_1^{++}$ plaquette operator
estimates $m(1)$ from correlations $0\le t\le 1$ and $m(2)$ from 
$1\le t\le 2$. In the disordered phase one has also signals for $t=3$ 
which tend to give somewhat lower values than $m(2)$, but are so noisy 
that they overlap within statistical errors with $m(2)$. In the ordered 
phase signals are even weaker, so that the correlations at $t=2$ 
include in most cases zero within two error bars. We give therefore 
only the $m(1)$ values, which lie considerably higher than in the 
disordered phase, with a hysteresis visible at the transition. One 
may conjecture that there are no glueball states in an eventual 
quantum continuum limit of the ordered phase.  Instead, as shown next, 
with vanishing SU(2) string tension the glueball spectrum appears to 
break up into massive vector bosons.

Within Higgs model simulations on the lattice trial operators 
for the $W$ mass are given by~\cite{MM1}
\begin{equation} \label{Wboson}
  W_{i,\mu}(x)\ =\ -i\,{\rm Tr}\,\left[\tau_i\,W_{\mu}(x)\right]\,,
\end{equation}
where $W_{\mu}(x)$ is the gauge invariant link variable
\begin{equation} \label{Wlink}
  W_{\mu}(x)\ =\ g^{\dagger}(x+\hat{\mu}a)\,V_{\mu}(x)\,g(x)
\end{equation}
and the SU(2) matrix $g(x)$ collects the angular variables of a 
complex doublet scalar field. This is also a gauge invariant operator
for our action (\ref{Sscalar}) with scalar fields. Our simulations 
correspond to this action at very large $\kappa$ after fixing the 
gauge, so that (\ref{Wboson}) becomes
\begin{equation} \label{Vboson}
  V_{i,\mu}(x)\ =\ -i\,{\rm Tr}\,\left[\tau_i\,V_{\mu}(x)\right]\,.
\end{equation}
In the following we calculate vector boson masses from correlations 
of this operator.

\begin{table}[tb]
\caption{Effective $0^{++}$ glueball masses for $t=1,\,2$ and vector
boson mass $m_W$. \label{tab_masses}} 
\centering
\begin{tabular}{|c|c|c|c|c|c|}  \hline
&\multicolumn{2}{c|}{$\lambda=0.4$}&\multicolumn{3}{c|}{$\lambda=0.9$}
\\ \hline 
Lat      & $m(1)$     & $m(2)$    & $m(1)$    & $m(2)$   & $m_W$     \\ \hline
$4^3 16$ &2.144~~(14) &1.551 (49) &3.471 (40) & noise    &0.4706 (65)\\ \hline
$6^3 24$ &2.0925 (46) &1.451 (23) &3.380 (21) &2.99 (32) &0.3385 (32)\\ \hline
$8^3 32$ &2.1175 (37) &1.509 (18) &3.396 (16) &2.70 (26) &0.2901 (13)\\ \hline
$10^348$ &2.1144 (38) &1.498 (13) &3.393 (12) &2.91 (24) &0.2729 (09)\\ \hline
$12^364$ &2.1173 (32) &1.509 (12) &3.400 (12) &2.83 (20) &0.2677 (09)\\ \hline
\end{tabular} \end{table} 

\begin{figure}[tb] \begin{center} 
\epsfig{figure=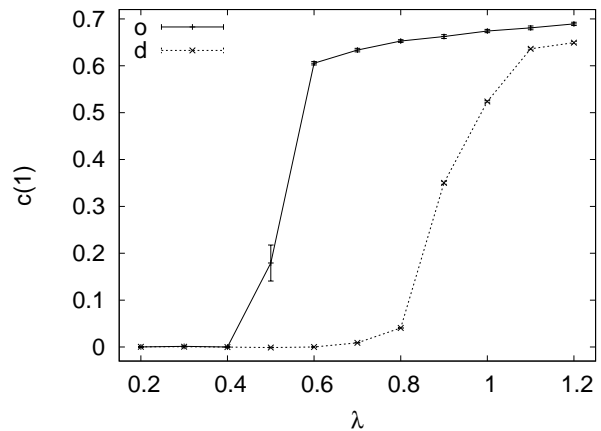,width=\columnwidth} 
\caption{Vector boson correlation function on a $4^316$ lattice at $t=1$ 
(o ordered and d disordered SU(2) starts).\vspace{-4mm}\label{fig_cor1}} 
\end{center} \end{figure}

In contrast to the zero momentum correlations of our glueball trial 
operators, those of $V_{i,\mu}$ turn out to be beautifully strong in 
the ordered phase, while they are zero in the disordered phase. For 
the $4^316$ lattice the hysteresis of the $t=1$ correlation is shown 
in Fig.~\ref{fig_cor1}. At large $\lambda$ values ($\lambda=1.0$ 
and $\lambda=1.2$) domain walls appear to prevent equilibration of 
disordered starts. Mass estimates in the ordered phase from cosh 
fits (\ref{coshE}) are found in the lower right part of 
Fig.~\ref{fig_masses} (left of the transition $m_W$ masses are 
infinite). As these masses rely on long range correlations finite 
size corrections are substantial. For several lattice sizes $m_W$ 
estimates at $\lambda =0.9$ are collected in table~\ref{tab_masses}, 
where we also see that there are almost no finite size effects for 
glueball masses.

\begin{figure}[tb] \begin{center} 
\epsfig{figure=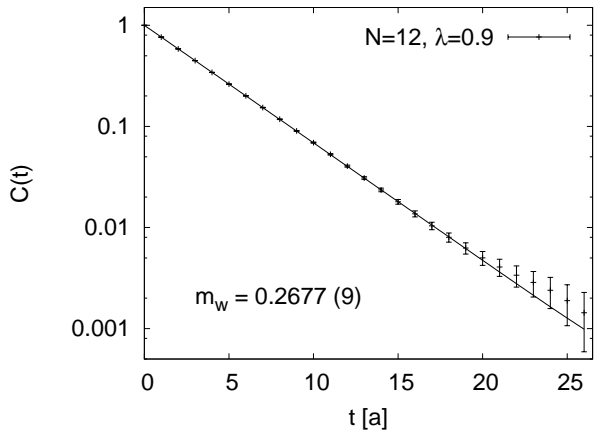,width=\columnwidth} 
\caption{Correlation function data and fit for $m_W$ mass on a
$12^3 64$ lattice. \vspace{-4mm} 
\label{fig_Wcor} } 
\end{center} \end{figure}

\begin{figure}[tb] \begin{center} 
\epsfig{figure=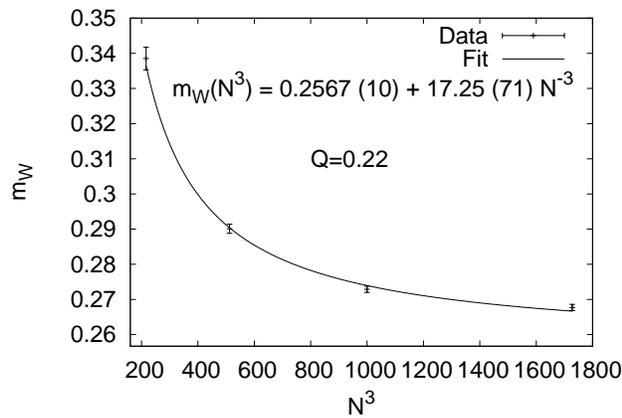,width=\columnwidth} 
\caption{Finite size extrapolation of the $W$ mass. \vspace{-4mm} 
\label{fig_Wfsize} } 
\end{center} \end{figure}

The correlation function for the $m_W$ fit at $\lambda=0.9$ on our 
largest lattice is depicted in Fig.~\ref{fig_Wcor}. The signal can be 
followed up to more than 20 lattice spacings in the $N_t$ extension. 
While significant for small volumes, finite size corrections decrease 
quickly for larger volumes. For our $N\ge 8$ lattices a fit of the 
form $m_W(N)=m_W+const/N^3$ works well and is depicted in 
Fig.~\ref{fig_Wfsize}. It yields the infinite volume 
extrapolation $m_W= 0.2567\,(10)$ with $Q=0.22$ goodness of fit.

\section{Summary, outlook and conclusions \label{sec_sum}}

We have introduced a model of SU(2) and U(1) vector fields, which
can be obtained from the gauge invariant interaction (\ref{Sscalar}) 
between a scalar matrix field and the vector fields in the London 
limit $(\Phi^{\dagger}\Phi)^2\to\tau_0$ ($\tau_0\ 2\times2$ unit 
matrix), i.e., $\kappa\to \infty$ in Eq.~(\ref{Sscalar}). On a 
finite lattice this interaction is for sufficiently large $\kappa$ 
supposed to be indistinguishable from the $\kappa\to\infty$ limit.
MC simulations of the quantum field theory on the lattice exhibits a 
phase transition between a deconfined phase with a glueball spectrum 
and a deconfined phase with a massive vector boson triplet. A massless
photon is found in the spectrum of both phases. 

Whether this quantum field theory just lives on the lattice or has a 
quantum continuum limit remains to be clarified. The question of its 
renormalizability requires perturbative investigations, which are 
beyond the scope of this paper. Due to the London limit the scalar
field becomes that of a non-linear $\sigma$ model, which is 
usually non-renormalizable. However, our situation is peculiar,
because the scalar field can be absorbed into extended gauge 
transformations of the SU(2) and U(1) vector fields (\ref{aGU}),
(\ref{aGV}) and the theory we are interested in is the one without
scalar fields as first formulated in~\cite{Be09a}.

In itself the lattice properties are rather remarkable, most 
of all the evidence from Fig.~\ref{fig_cor1} to~\ref{fig_Wfsize} 
for a massive vector boson triplet in the deconfined, but not in 
the confined, phase.

\acknowledgments 
This work was in part supported by the DOE grant DE-FG02-97ER41022 
and by a Research Award of the Humboldt Foundation. Part of the work 
was done at Leipzig University and I am indebted to Wolfhard Janke and 
his group for their kind hospitality. Further, I thank Holger Perlt and 
Arwed Schiller for useful discussions. Some of the computer programs 
used rely on collaborations with Alexei Bazavov and benefited from 
programing help by Hao Wu.

\appendix 

\section{Invariance under Extended Gauge Transformations. \label{appA}}

Following \cite{Be09b} we show the invariance of (\ref{Lint}) under 
extended gauge transformations.  Expanding a calculation of \cite{Qu83} 
slightly, we find
\begin{eqnarray} 
  &~& g_a\partial_{\mu}A'_{\nu} - g_b\partial_{\nu}B'_{\mu} = 
  \\ \nonumber
  &~& \partial_{\mu}[Gg_aA_{\nu}G^{-1}+i(\partial_{\nu}G)G^{-1}] 
  \\ \nonumber 
  &-& \partial_{\nu}[Gg_bB_{\mu}G^{-1}+i(\partial_{\mu}G)G^{-1}]
  =\\ \nonumber &~& \\ 
  &~& G(g_a\partial_{\mu}A_{\nu} - g_b\partial_{\nu}B_{\mu})G^{-1}
  \\ \nonumber 
  &+& [(\partial_{\mu}G)g_aA_{\nu}-(\partial_{\nu}G)g_bB_{\mu}]G^{-1}
  \\ \nonumber &+& 
  G[g_aA_{\nu}(\partial_{\mu}G^{-1})-g_bB_{\mu}(\partial_{\nu}G^{-1})]
  \\ \nonumber &+& 
  i\,[(\partial_{\nu}G)(\partial_{\mu}G^{-1})-
      (\partial_{\mu}G)(\partial_{\nu}G^{-1})]
\end{eqnarray}
Using $(\partial_{\mu}G^{-1})G+G^{-1}(\partial_{\mu}G)=\partial_{\mu}
(G^{-1}G)=0$, this can be transformed to
\begin{eqnarray} \label{dApdBp}
  &~&g_a\partial_{\mu}A'_{\nu} - g_b\partial_{\nu}B'_{\mu} =\\ \nonumber
  &~& G\left(g_a\partial_{\mu}A_{\nu} - g_b\partial_{\nu}B_{\mu}
       \right)G^{-1} \\ \nonumber &+&
  G\left\{\left[G^{-1}(\partial_{\mu}G),g_aA_{\nu}\right] -
  \left[G^{-1}(\partial_{\nu}G),g_bB_{\mu}\right]\right\}G^{-1}
  \\ \nonumber 
  &-& i\,G[(\partial_{\mu}G^{-1})(\partial_{\nu}G) -
           (\partial_{\nu}G^{-1})(\partial_{\mu}G)]G^{-1}\,.
\end{eqnarray}
The commutator term transforms as
\begin{eqnarray} 
  i\,g_ag_b&&\!\!\!\!\!\![B'_{\mu},A'_{\nu}] = \\ \nonumber
  i\,g_ag_b&&\!\!\!\!\!\![(GB_{\mu}G^{-1}+(i/g_b)(\partial_{\mu}G)G^{-1}),
  \\ \nonumber &&\!\!\!\!(GA_{\nu}G^{-1}+(i/g_a)(\partial_{\nu}G)G^{-1})]
\end{eqnarray}
\begin{eqnarray} \label{commutator}
  &=& i\,g_ag_b G[B_{\mu},A_{\nu}]G^{-1} \\ \nonumber
  &-& G\{[G^{-1}(\partial_{\mu}G),g_aA_{\nu}] -
         [G^{-1}(\partial_{\nu}G),g_bB_{\mu}]\}G^{-1} \\ \nonumber
  &+& i\,G[(\partial_{\mu}G^{-1})(\partial_{\nu}G) -
           (\partial_{\nu}G^{-1})(\partial_{\mu}G)]G^{-1}\,.
\end{eqnarray}
Combining (\ref{dApdBp}) and  (\ref{commutator}) yields (\ref{Fintg}).

\section{Classical continuum limit of the action with scalar fields.
\label{appB}}

We write the scalar matrix field of the action (\ref{Sscalar}) as
\begin{equation} \label{ph1ph1}
  \Phi(x)\ =\ \phi_1(x)\,\phi_2(x)
\end{equation}
with $\phi_1$ and $\phi_2$ defined by (\ref{phi1phi2}) and factor
(\ref{Sscalar}) into the form
\begin{eqnarray} \nonumber
  S_{\mu\nu}^s &=& \frac{\lambda^s}{4}\,{\rm Re\, Tr}\,\left(
  S_{1\mu\nu}\right)\,{\rm Tr}\,\left(S_{2\mu\nu}\right) + h.c. \\
  \label{S1S2} &+& \kappa\,{\rm Tr}\,[(\Phi^{\dagger}\Phi-\tau_0)^2]\,.
\end{eqnarray}
As calculation of the classical continuum limit of this action leads 
to tedious algebra, we rely on the algebraic program FORM~\cite{Form}.
For this it is convenient to write (\ref{S1S2}) in a symmetric form 
with the position $x$ at the center of the plaquette:
\begin{eqnarray}
  && S_{1\mu\nu} = \\ \nonumber && \left[
  \phi_1^{\dagger}\left(x-\hat{\mu}\frac{a}{a}-\hat{\nu}\frac{a}{2}\right)
  \,U_{\mu}\left(x-\hat{\nu}\frac{a}{2}\right)\,
  \phi_1\left(x+\hat{\mu}\frac{a}{2}-\hat{\nu}\frac{a}{2}\right)\right]
  \\ \nonumber && \left[
  \phi_1^{\dagger}\left(x-\hat{\mu}\frac{a}{2}+\hat{\nu}\frac{a}{2}\right)
  \,U_{\mu}\left(x+\hat{\nu}\frac{a}{2}\right)\,
  \phi_1\left(x+\hat{\mu}\frac{a}{2}+\hat{\nu}\frac{a}{2}\right)
  \right]^{\dagger}\!\!\!, \\
   && \!\! S_{2\mu\nu} = \\ \nonumber &&\!\! \left[
  \phi_2^{\dagger}\left(x+\hat{\mu}\frac{a}{2}-\hat{\nu}\frac{a}{2}\right)
  \,V_{\nu}\left(x+\hat{\mu}\frac{a}{2}\right)
  \,\phi_2\left(x+\hat{\mu}\frac{a}{2}+\hat{\nu}\frac{a}{2}\right)\right]
  \\ && \!\! \nonumber \left[
  \phi_2^{\dagger}\left(x-\hat{\mu}\frac{a}{2}-\hat{\nu}\frac{a}{2}\right)
  \,V_{\nu}\left(x-\hat{\mu}\frac{a}{2}\right)
  \,\phi_2\left(x-\hat{\mu}\frac{a}{2}+\hat{\nu}\frac{a}{2}\right)
  \right]^{\dagger}\!\!\! .
\end{eqnarray}
We expand the scalar fields $\phi_i$, $i=1,2$, to order $a^2$. With 
$\epsilon$ and $\eta$ of order $a$: 
\begin{eqnarray} \nonumber
  \phi_i(x+\epsilon\hat{\mu}+\eta\hat{\nu}) &=& \phi_i(x)+
  \epsilon\partial_{\mu}\phi_i(x) + \eta\partial_{\nu}\phi_i(x)
  \\ \label{phiTaylor} &+& 
  \eta\epsilon\partial_{\nu}\partial_{\mu}\phi_i(x) + \dots~.
\end{eqnarray}
The gauge matrices defined by (\ref{UV}) have to be expanded up to 
order $a^4$ in the lattice spacing. For the gauge potential the
substitutions
\begin{eqnarray} 
  A_{\mu}\left(x+\eta\hat{\nu}\right) &=&
  A_{\mu}(x) + \eta\partial_{\nu}A_{\mu}(x)\,,\\
  B_{\nu}\left(x+\epsilon\hat{\mu}\right) &=&
  B_{\nu}(x) + \epsilon\partial_{\mu}B_{\nu}(x)\,,
\end{eqnarray}
are done. These expansions generate more terms (the computer does not 
care) than the analogue expansions of (\ref{Sscalar}), but avoid some 
complications in the identification of covariant operators. We write 
now for $i=1,2$
\begin{equation} \label{SiExpansion}
  S_{i\mu\nu} = \left(\phi_i^{\dagger}\phi_i\right)^2 + 
  a^2\,S_{i\mu\nu}(2) + a^4\,S_i(4) + \dots~.
\end{equation} \smallskip
The traces of the order $a$ and $a^3$ terms of this expansion are seen 
to vanish. The covariant contribution of
\begin{equation} \label{ReTrS1S2}
  {\rm Re}\,{\rm Tr}\,\left[S_{1\mu\nu}(2)\right]\,{\rm Tr}\,
  \left[S_{2\mu\nu}(2)\right]\,
\end{equation}
to (\ref{S1S2}) comes from the anti-Hermitean terms
\begin{eqnarray} \nonumber &&
  {\rm Tr}\,\left[S_{2\mu\nu}(2)\right] = \frac{1}{4}\,{\rm Tr}\,
  \left\{ \left(D^b_{\nu}\,\phi_2\right)^{\dagger}\phi_2\,
          \left(D^b_{\mu}\,\phi_2\right)^{\dagger}\phi_2\,
  \right.  \\ \nonumber && 
      + \left(D^b_{\nu}\,\phi_2\right)^{\dagger} \phi_2\
        \phi_2^{\dagger}\,D^b_{\mu}\,\phi_2 +
        \left(D^b_{\mu}\,\phi_2\right)^{\dagger} \phi_2\,
        \left(D^b_{\nu}\,\phi_2\right)^{\dagger} \phi_2 
  \\ \nonumber && 
      - \left(D^b_{\mu}\,\phi_2\right)^{\dagger} \phi_2\
        \phi_2^{\dagger}\,D^b_{\nu}\,\phi_2 -
        \phi_2^{\dagger}\,D^b_{\nu}\,\phi_2\,
        \left(D^b_{\mu}\,\phi_2\right)^{\dagger} \phi_2 
  \\ \nonumber && 
      - \phi_2^{\dagger}\,D^b_{\nu}\,\phi_2\,
        \phi_2^{\dagger}\,D^b_{\mu}\,\phi_2 +
        \phi_2^{\dagger}\,D^b_{\mu}\,\phi_2\,
        \left(D^b_{\nu}\,\phi_2\right)^{\dagger} \phi_2 
  \\ \nonumber && 
      - \phi_2^{\dagger}\,D^b_{\mu}\,\phi_2\,
        \phi_2^{\dagger}\,D^b_{\nu}\,\phi_2 -
        2\,\phi_2^{\dagger}\,\phi_2\,\left(D^b_{\nu}\,
        \phi_2\right)^{\dagger}D^b_{\mu}\,\phi_2 
  \\ \nonumber && 
      - 2\,\phi_2^{\dagger}\,\phi_2\,\left(D^b_{\mu}\,
        D^b_{\nu}\,\phi_2\right)^{\dagger}\,\phi_2 +
        2\,\phi_2^{\dagger}\,\phi_2\,\left(D^b_{\mu}\,
        \phi_2\right)^{\dagger}D^b_{\nu}\,\phi_2 
  \\ &&\left.
      + 2\,\phi_2^{\dagger}\,\phi_2\,\phi_2^{\dagger}
        D^b_{\mu}\,D^b_{\nu}\,\phi_2 \right\} \label{TrO22}
\end{eqnarray}
and ${\rm Tr}\,\left[S_{1\mu\nu}(2)\right]$, which is obtained by 
interchanging $A_{\alpha}$ with $B_{\alpha}$ ($D^a_{\alpha}$ with 
$D^b_{\alpha}$) and then the subscripts $\mu$ and $\nu$. 

In the London limit ${\rm Tr}\,\left[S_{1\mu\nu}(2)\right]\,{\rm Tr}\,
\left[S_{2\mu\nu}(2)\right]$ gives the contribution $\sim 
{\rm Tr}\,\left(\partial_{\nu}A_{\mu}\,\partial_{\mu}B_{\nu}\right)$ 
to the action~(\ref{Lint}). The terms $\sim {\rm Tr}\,\left(
\partial_{\nu}A_{\mu}\,\partial_{\nu}A_{\mu}\right)$ and $\sim {\rm Tr}
\,\left(\partial_{\mu}B_{\nu}\,\partial_{\mu}B_{\nu} \right)$ of 
(\ref{Lint}) come from the $a^4\,(\phi_1^{\dagger}\,\phi_1)^2\,
S_{2\mu\nu}(4)$ and $a^4\,S_{1\mu\nu}(4)\,(\phi_2^{\dagger}\,\phi_2)^2$ 
contributions to (\ref{S1S2}), whose calculation we found too tedious 
to pursue. For small oscillations about the expectation value of the 
scalar fields 
\begin{equation} 
  S_{1\mu\nu}(4) = S_{1\mu\nu}(2)^2~~{\rm and}~~
  S_{2\mu\nu}(4) = S_{2\mu\nu}(2)^2
\end{equation}
are good approximations, illustrating the typical gauge invariant 
terms encountered.
\hfill\break 


\begin{thebibliography}{19}
\bibitem{MMH} Compare chapter~6.1.1 of~\cite{MM94} ($\kappa$ and
              $\lambda$ are interchanged with respect to their
              notation).
\bibitem{MM1} Eq.~(6.27) of~\cite{MM94} (their $V$ is our $W$ and 
              our $V$ their $U$).
\end{thebibliography}
\end{document}